\documentclass[sigconf,nonacm]{acmart}

\setcopyright{none}
\renewcommand\footnotetextcopyrightpermission[1]{}

\usepackage{amsmath,amsfonts}
\usepackage{array}
\usepackage{booktabs}
\usepackage{dblfloatfix}
\usepackage{boxedminipage}
\usepackage{bbding}
\usepackage{enumitem}
\usepackage{framed}
\usepackage{graphicx}
\usepackage{listings}
\usepackage{makecell}
\usepackage{mdframed}
\usepackage{multirow}
\usepackage{pifont}
\usepackage[caption=false,font=footnotesize]{subfig}
\usepackage{tabularx}
\usepackage{threeparttable}
\usepackage[many]{tcolorbox}
\usepackage{xspace}

\newcommand{\tool}{\emph{ReSource}\xspace}
\newcommand{\degpt}{\emph{DeGPT}\xspace}
\newcommand{\fide}{\emph{FidelityGPT}\xspace}
\newcommand{\llm}{\emph{LLM4Decompile}\xspace}
\newcommand{\ReSourceg}{\tool$_{\text{w/o CFS}}$\xspace}
\newcommand{\ReSourcec}{\tool$_{\text{w/o GDS}}$\xspace}
\newcommand{\ReSourcef}{\tool$_{\text{w/o G+C}}$\xspace}

\begin{document}

\title{Decoupling is a Necessity: Transformation-Agnostic Decompiled Code Recovery under Optimization and Obfuscation}

\author{Zhiping Zhou}
\affiliation{\institution{Tianjin University}\city{Tianjin}\country{China}}
\email{zhou_zhiping@tju.edu.cn}

\author{Xiaohong Li}
\affiliation{\institution{Tianjin University}\city{Tianjin}\country{China}}
\email{xiaohongli@tju.edu.cn}

\author{Ruitao Feng}
\authornote{Joint corresponding authors.}
\affiliation{\institution{Southern Cross University}\city{Lismore}\country{Australia}}
\email{ruitao.feng@scu.edu.au}

\author{Yao Zhang}
\authornotemark[1]
\affiliation{\institution{Tianjin University}\city{Tianjin}\country{China}}
\email{zzyy@tju.edu.cn}

\author{Yuekang Li}
\affiliation{\institution{University of New South Wales}\city{Sydney}\country{Australia}}
\email{yuekang.li@unsw.edu.au}

\author{Wenbu Feng}
\affiliation{\institution{Tianjin University}\city{Tianjin}\country{China}}
\email{ianx@tju.edu.cn}

\renewcommand{\shortauthors}{Zhou et al.}
\hypersetup{pdfauthor={Zhiping Zhou; Xiaohong Li; Ruitao Feng; Yao Zhang; Yuekang Li; Wenbu Feng}}

\begin{abstract}
Reverse engineering is essential for software security analysis and vulnerability detection. Decompilation, the process of lifting binaries to high-level pseudocode, is central to this task. However, production binaries are hostile environments: aggressive compiler optimizations and adversarial obfuscation jointly mangle control structures, obscure variable intents, and disguise high-level program logic. Consequently, existing LLM-based decompilation tools frequently suffer from structural collapse and semantic hallucinations. We present \tool, the first multi-phase LLM framework designed for transformation-agnostic source recovery. To tackle these intertwined distortions, \tool conceptualizes the binary-to-source discrepancies into three orthogonal tiers, namely lexical, syntactic, and semantic, and decouples the recovery process accordingly. First, to ground the LLM and prevent logic drift, it retrieves empirical priors from a curated Semantic Distortion Database. Second, to resolve control-flow flattening, it integrates a lightweight predictor to reconstruct the source-level structural skeleton. Finally, a contextual lexical deduction stage refines identifiers to restore human readability. Evaluated on a massive benchmark of over 80,000 decompiled--source function pairs across three optimization levels and four obfuscation techniques, \tool achieves an 83\% Top-5 source retrieval accuracy and an average similarity score of 0.66. By maintaining robust semantic identifiability where state-of-the-art baselines (\degpt, \llm, and \fide) severely overfit or degrade, \tool provides a scalable and reliable foundation for downstream security analysis.
\end{abstract}

\ccsdesc[500]{Security and privacy~Software reverse engineering}
\keywords{Reverse Engineering, Decompilation, Compiler Optimizations, Code Obfuscation, Large Language Models}

\maketitle
\hypersetup{pdfauthor={Zhiping Zhou; Xiaohong Li; Ruitao Feng; Yao Zhang; Yuekang Li; Wenbu Feng}}

\section{Introduction}
\label{sec:Introduction}

Reverse engineering~\cite{muller2000reverse} is the cornerstone of software security analysis, enabling critical downstream tasks like patch diffing~\cite{xu2017spain,zhao2020patchscope}, binary similarity analysis~\cite{hu2017binary,luo2017semantics}, and vulnerability discovery. Because analyzing low-level assembly or graphs~\cite{gu2025uniasm,jiang2020similarity,luo2023vulhawk,zhang2024gtrans} struggles to support source-level reasoning, analysts heavily rely on decompilers (e.g., IDA Pro~\cite{hex-rays}, Ghidra~\cite{ghidra}) to lift binaries into C-like pseudocode. Recently, the structural similarity between this pseudocode and natural programming languages has prompted a surge in adopting Large Language Models (LLMs) to bridge the semantic gap between machine-generated artifacts and human-readable source code.

Current LLM-assisted decompilation generally follows two trajectories. The first is direct translation, such as fine-tuning models on massive assembly-to-C datasets~\cite{tan2024llm4decompile}. The second is post-processing, commonly defined in this domain as \emph{source recovery}, which involves taking raw decompiler pseudocode and restoring its original structural and semantic intent to produce readable, high-fidelity source representations~\cite{hu2024degpt,zhou2025fidelitygpt}.

Despite these advances, existing frameworks operate under a critical flaw: they primarily target benign, lightly optimized code (e.g., O1 or O2). In reality, production binaries are hostile environments, heavily transformed by aggressive compiler optimizations (e.g., O3) and adversarial obfuscation (e.g., Control-Flow Flattening). Because current LLM approaches rely heavily on superficial sequence-to-sequence pattern matching, they severely overfit to benign transformation distributions. When confronted with the intertwined chaos of optimization and obfuscation, these models suffer catastrophic structural collapse and hallucinate incorrect logic, yielding code that is semantically hazardous for downstream security tasks~\cite{zhang2025unseen}.

To bridge this critical gap and enable faithful recovery from heavily optimized and obfuscated binaries, we identify four core challenges that must be systematically addressed:

\textbf{C1: Unified recovery across diverse transformations.} Since the exact combination of optimizations and obfuscations is typically unknown during analysis, recovery methods must avoid overfitting to specific patterns and instead handle heterogeneous distortions through a unified formulation.

\textbf{C2: Preserving semantic fidelity against opaque artifacts.} Aggressive transformations replace human-readable logic with opaque artifacts. Confronted with these alien patterns, vanilla LLMs lose their reasoning context and frequently hallucinate incorrect business logic.

\textbf{C3: Structural collapse from control-flow distortion.} Adversarial transformations like control-flow flattening completely obliterate program hierarchies, dismantling readable loops and branches into artificial state machines. Standard sequence-to-sequence LLMs struggle to naturally ``un-flatten'' this structural chaos.

\textbf{C4: Limitations of superficial evaluation.} Traditional metrics (e.g., BLEU) reward superficial token overlap but are blind to functional equivalence under structural divergence. Robust assessment must directly measure semantic identifiability to ensure real-world utility.

To tackle these challenges, we reframe decompilation: faithful recovery is not a monolithic sequence-to-sequence mapping, but a multi-dimensional reconstruction problem requiring explicit structural and empirical priors. Guided by this principle, we present \tool, the first multi-phase LLM framework for transformation-agnostic source recovery. To achieve unified recovery (addressing C1), \tool decouples the process across lexical, syntactic, and semantic tiers. First, it queries a novel Semantic Distortion Database to retrieve empirical priors, resolving opaque artifacts and mitigating hallucinations (addressing C2). Second, a lightweight control-flow predictor reconstructs the source-level architectural skeleton to reverse structural collapse (addressing C3). Finally, lexical finalization refines identifiers to restore human readability.

We evaluate \tool on a massive benchmark of over 80,000 function pairs across three optimization levels and four obfuscation schemes. To directly validate semantic identifiability and overcome the limits of shallow evaluation (addressing C4), we introduce a rigorous retrieval-based methodology. Experimental results show that \tool consistently outperforms state-of-the-art baselines (\degpt, \fide, \llm), achieving an 83\% average Top-5 retrieval accuracy and maintaining remarkable stability even under severe control-flow obfuscation.

\textbf{Contributions.} This paper makes the following key contributions:
\begin{itemize}[leftmargin=*]
  \item We propose \tool, the first multi-stage LLM framework for transformation-agnostic source recovery against the joint threats of aggressive optimization and adversarial obfuscation.
  \item We design a synergistic architecture that decouples semantic guidance (via a data-driven Semantic Distortion Database) from structural repair (via a skeleton predictor), effectively neutralizing LLM hallucinations and structural collapse.
  \item We construct a comprehensive benchmark of 80,000+ binary-source pairs and introduce a rigorous retrieval-based evaluation methodology to accurately assess global semantic identifiability.
\end{itemize}

\section{Background \& Motivation}
\label{sec:Background}

We first review the inherent limitations of current decompilation and LLM-based recovery techniques~(\S\ref{sec:background}). We then dissect a real-world obfuscation scenario~(\S\ref{sec:motivation}) to expose the core challenges that motivate the design of \tool.

\subsection{Background}
\label{sec:background}

\textbf{Decompilation under Aggressive Transformations.} Modern compilers employ optimizations~\cite{kandemir2000influence} and obfuscations~\cite{schloegel2022loki} that fundamentally restructure binaries, creating a severe ``semantic gap''~\cite{liu2020far,yakdan2015no}. Transformations like control-flow flattening discard high-level abstractions, leaving decompilers to produce fragmented pseudocode with mangled logic. This loss of structural and lexical fidelity severely hinders downstream security tasks.

\textbf{Limitations of LLM-based Recovery.} While LLMs~\cite{zhao2023survey} excel at code summarization, they struggle to generalize across diverse compiler-induced distortions~\cite{tan2024llm4decompile,xie2024resym,xu2025unleashing}. Purely sequence-to-sequence approaches fail to restore deep structural logic or provide interpretable links back to source semantics under aggressive transformations. This necessitates a hybrid approach that explicitly grounds LLM reasoning with semantic and structural priors.

\subsection{Motivation}
\label{sec:motivation}

\begin{figure*}[t]
  \centering
  \includegraphics[width=\textwidth]{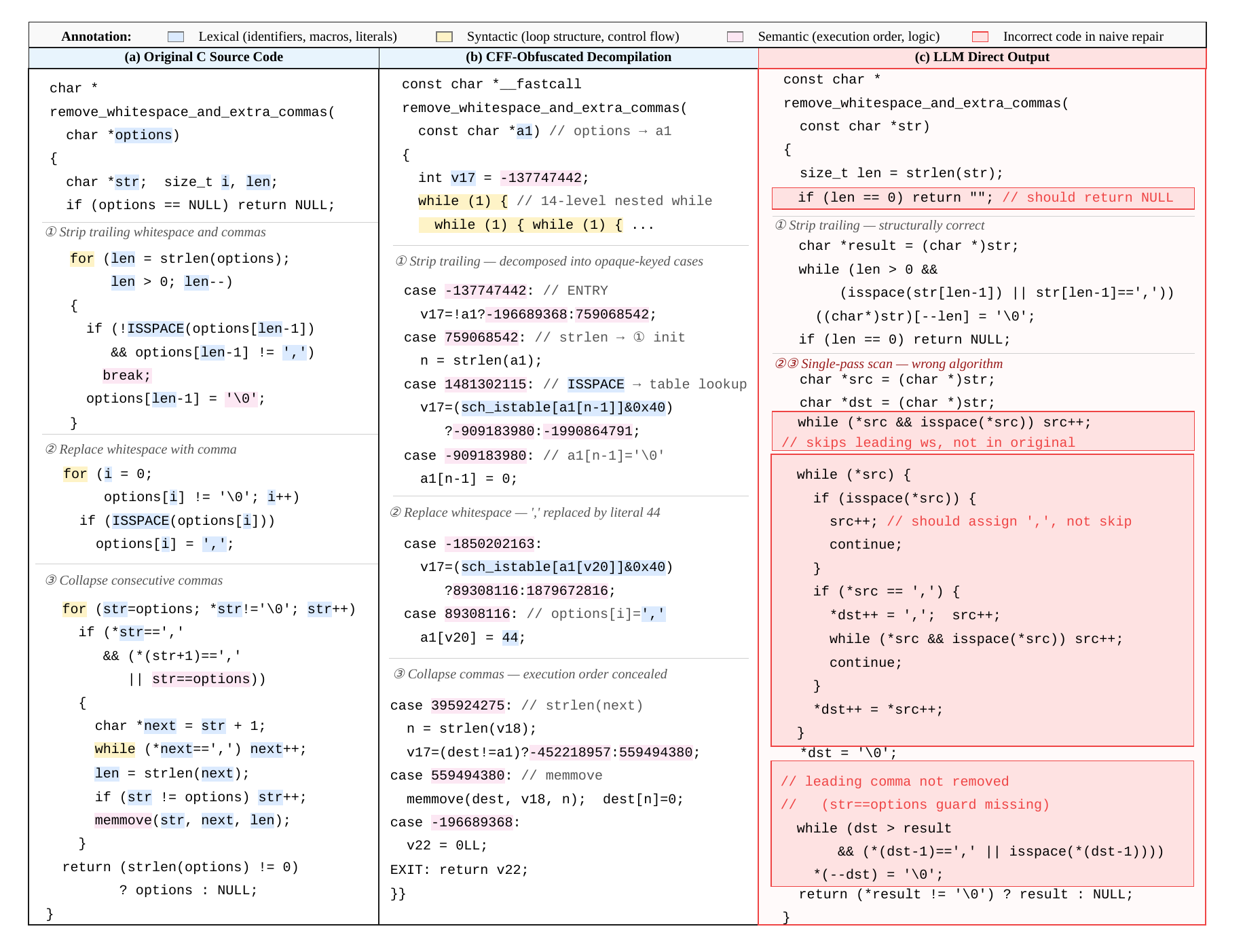}
  \caption{Motivating Example}
  \label{fig:motivating_example}
\end{figure*}

We illustrate the inherent limitations of pure sequence-to-sequence LLMs through a real-world obfuscation scenario. Figure~\ref{fig:motivating_example} presents (a) the original source code, alongside (b) its decompiled counterpart after Control-Flow Flattening (CFF), and (c) a failed direct-recovery attempt by a vanilla LLM.

\textbf{Challenge 1: Multi-Level Discrepancies.} As illustrated in Figure~\ref{fig:motivating_example}(b), adversarial transformations like CFF introduce massive visual and logical divergence between the source and the decompiled code. To decouple the recovery process from specific obfuscators, we abstract these massive discrepancies into three unified tiers:
\begin{itemize}[leftmargin=*]
  \item \textbf{Lexical:} Human-readable cues evaporate. The \texttt{ISSPACE} macro is forcefully inlined into cryptic bitwise operations (\nolinkurl{sch_istable[a1[n-1]]&0x40}), and literals are replaced by ASCII values (44).
  \item \textbf{Syntactic:} High-level constructs are rewritten. The source's clean, sequential \texttt{for} loops are entirely flattened into a monolithic switch-case block nested within a 14-level deep \texttt{while(1)} loop.
  \item \textbf{Semantic:} The driving force of the logic diverges from original intent. Data-driven control flow is hijacked by an artificial state machine driven by opaque assignments (e.g., \texttt{v17 = -137747442}).
\end{itemize}

\textbf{Challenge 2: Hallucinations from Missing Semantic Guidance.} When confronted with opaque decompiler artifacts and stripped identifiers, pure sequence-to-sequence LLMs lack the contextual priors necessary to infer original intent. Without meaningful guidance, the model resorts to unreliable guesswork. As shown in Figure~\ref{fig:motivating_example}(c), this lack of context causes the LLM to hallucinate incorrect business logic: it invents a non-existent feature to strip leading whitespaces, incorrectly skips characters instead of replacing them, and misses crucial boundary guards. Preventing this requires injecting concrete source-level suggestions to explicitly ground generation.

\textbf{Challenge 3: Structural Recovery Limitations.} Figure~\ref{fig:motivating_example}(b) illustrates how CFF obliterates conventional control-flow structure. Mapping sequences token-by-token without global structural priors, conventional LLMs fail to ``un-flatten'' such disordered code. When tasked to directly deobfuscate (Figure~\ref{fig:motivating_example}(c)), the LLM is overwhelmed by structural chaos. It abandons the original architecture, severely restructuring the algorithm into a monolithic, buggy pass. To address this, \tool integrates a prediction module that explicitly infers the source-level control skeleton before generation.

\textbf{Challenge 4: Limitations of Shallow Evaluation.} Aggressive transformations render traditional metrics like BLEU or token overlap~\cite{geng2024large,papineni2002bleu} insufficient. They fail to capture functional equivalence under structural divergence and may incorrectly reward hallucinated code. To address this, we move beyond shallow textual similarity and adopt a retrieval-based evaluation (i.e., identifying the exact source from a massive corpus). This robust proxy for semantic fidelity reflects the code's true utility in real-world security tasks.

\textbf{Bridging the Gap with \tool.} Driven by these four challenges, we design \tool, a unified pipeline bridging the binary-to-source gap through three targeted stages: (1) \textbf{Semantic Guidance} (\S\ref{sec:semantic-guidance}): retrieving context-aware suggestions to resolve multi-level ambiguities; (2) \textbf{Structural Repair} (\S\ref{sec:structure-repair}): predicting the original control-flow skeleton to fix flattening; and (3) \textbf{Lexical Finalization} (\S\ref{sec:finalization}): refining identifiers for human-readable clarity.

\section{Methodology}
\label{sec:methodology}

\subsection{Overview}

\begin{figure*}[t]
  \centering
  \includegraphics[width=\textwidth]{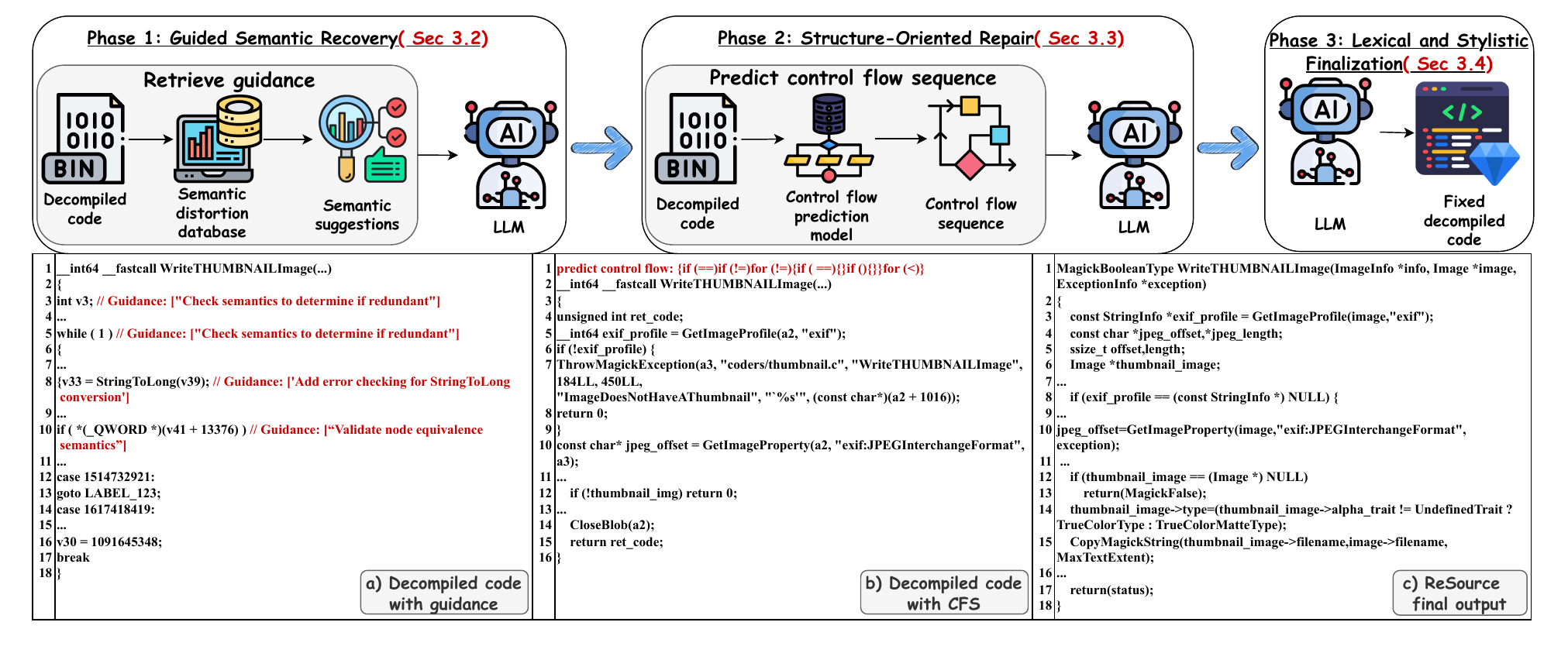}
  \caption{Overview of \tool}
  \label{fig:workflow}
\end{figure*}

Figure~\ref{fig:workflow} illustrates the architecture and progressive workflow of \tool. To systematically bridge the multi-dimensional discrepancies introduced by hostile transformations, \tool decouples the decompilation process into three synergistic phases:

\textbf{Phase 1: Guided Semantic Recovery (\S\ref{sec:semantic-guidance}).} \tool retrieves context-aware suggestions from a curated Semantic Distortion Database. These suggestions serve as empirical priors, constraining the LLM's reasoning toward the original developer intent and preventing logical hallucinations.

\textbf{Phase 2: Structure-Oriented Repair (\S\ref{sec:structure-repair}).} To reverse structural collapse (e.g., control-flow flattening), \tool predicts a source-level control skeleton. This structural prior guides the LLM to synthesize code with coherent loops and branches.

\textbf{Phase 3: Lexical Finalization (\S\ref{sec:finalization}).} The final stage refines variable identifiers and formatting to ensure human-readability and contextual consistency.

Each phase builds upon the previous, progressively injecting semantic, structural, and lexical constraints to achieve transformation-agnostic source recovery.

\subsection{Guided Semantic Recovery via Distortion Pattern Retrieval}
\label{sec:semantic-guidance}

Aggressive transformations replace human-readable logic with alien, compiler-induced artifacts (e.g., inlined macros, opaque bitwise math, or redundant state checks). When vanilla LLMs encounter these unfamiliar patterns, they lack the context to infer the original intent and frequently resort to semantic hallucinations. To break this cycle of guesswork, we shift the paradigm from unconstrained generation to retrieval-augmented deduction. We introduce the \textbf{Semantic Distortion Database}, a data-driven knowledge base that maps common decompiler artifacts to their ground-truth source-level intents.

\textbf{Core Insight: Leveraging the ``Seen'' to Decode the ``Unseen''.} Our approach is driven by a fundamental observation: while compilers and obfuscators generate seemingly infinite structural variations, the underlying semantic mappings between source idioms and resulting artifacts are highly recurrent across large codebases. Consequently, if an unseen decompiled snippet shares a similar semantic profile with seen historical artifacts, they almost certainly trace back to the same developer intent. By abstracting these reusable distortion patterns from a massive corpus of aligned code pairs, \tool retrieves concrete, historical recovery suggestions to hint the LLM. This effectively transforms the LLM's task from a blind translation into an informed repair process, using empirical priors to explicitly ground the recovery of unseen code.

\subsubsection{Constructing the Semantic Distortion Database}

\begin{figure*}[t]
  \centering
  \includegraphics[width=\textwidth]{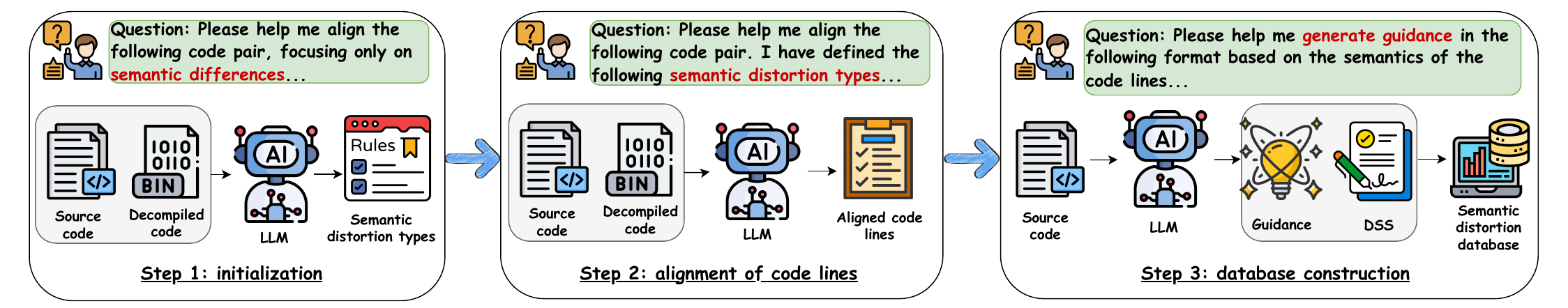}
  \caption{Construction Workflow of the Semantic Distortion Database}
  \label{fig:semantic_database}
\end{figure*}

Figure~\ref{fig:semantic_database} depicts the three-stage pipeline for constructing the database: taxonomy initialization, code alignment, and signature construction.

\textbf{Step 1: Distortion Taxonomy Initialization.} Understanding the nature of binary-to-source discrepancies is critical. Unlike prior work that manually defines fidelity categories based on expert intuition~\cite{dramko2024taxonomy,zhou2025fidelitygpt}, we adopt a data-driven approach. We prompt an LLM to identify and summarize actual semantic differences observed across massive aligned function pairs. After manual refinement and consolidation, we establish a compact yet expressive taxonomy of common semantic distortions:
\begin{itemize}[leftmargin=*]
  \item \textbf{Semantic Return:} Premature returns or incorrect return values.
  \item \textbf{Error Safety:} Missing or transformed exception-handling logic.
  \item \textbf{Condition Check:} Inverted, oversimplified, or altered conditional predicates.
  \item \textbf{Control Flow Transformation:} Flattened or irregular branch and loop structures.
  \item \textbf{State Management:} Divergent initialization or mutation of state variables.
  \item \textbf{Optimization Redundancy:} Redundant computations introduced by compilers.
  \item \textbf{Function Call:} Dropped return values, modified parameters, or reordered side effects.
  \item \textbf{Memory Usage:} Altered allocation, dereferencing, or memory access patterns.
\end{itemize}
This taxonomy underpins the database structure, organizing distortion patterns into semantically coherent categories for efficient retrieval.

\textbf{Step 2: Line-Level Code Alignment.} For each function in our corpus, we utilize LLM-assisted matching to align decompiled lines with their corresponding source-level counterparts. This yields a structured mapping comprising the \nolinkurl{decompiled_code_line} (with its line number), the matched \nolinkurl{source_code_line} (or null for compiler-induced artifacts), and the assigned \nolinkurl{distortion_type}. These aligned examples form the ground truth pairs for inference-time retrieval.

\textbf{Step 3: Guidance Generation and Semantic Signature Construction.} To support scalable recovery, we prompt an LLM to summarize the intended semantics of each \texttt{source\_code\_line} into natural language guidance. These guidance sentences serve as human-aligned recovery targets.

However, relying solely on raw lexical similarity (e.g., embedding distance of decompiled text) is insufficient for accurate retrieval. Decompiled code is notoriously stripped of contextual cues: for instance, the isolated statement \texttt{v1 = 0;} could signify the assignment of a numeric zero, a null pointer, a boolean false, an ASCII string terminator, or a purely redundant operation. Its true meaning depends entirely on its surrounding context.

To resolve such ambiguity, we construct a \textbf{Decompilation Semantic Signature (DSS)} for each distorted line. The DSS encodes semantics across four orthogonal dimensions:
\begin{itemize}[leftmargin=*]
  \item \textbf{Syntax Features:} Abstract code shape and AST-derived constructs.
  \item \textbf{Variable Dependencies:} Referencing context and data-flow usage.
  \item \textbf{Control Flow:} Relative nesting and control context within the function.
  \item \textbf{Contextual Role:} High-level behavioral role, such as condition guarding or computation.
\end{itemize}
Collectively, these components form a compact semantic sketch that enables fine-grained comparison across structurally diverse code snippets. The DSS serves as the indexing key for retrieving guidance during inference, ensuring that suggestions are matched contextually rather than just lexically.

\subsubsection{Semantic Suggestion Retrieval via Distortion-Aware Semantic Matching}

During inference, \tool retrieves context-aware semantic suggestions by matching distorted decompiled lines against annotated entries in the Semantic Distortion Database (Phase 1 of Figure~\ref{fig:workflow}). To balance retrieval efficiency and semantic precision without exceeding the LLM's context limits, we design a multi-stage retrieval and allocation strategy:

\textbf{Stage 1: Coarse Lexical Filtering.} Each unannotated decompiled line is first embedded using a pretrained dense encoder. To efficiently narrow the massive search space, we perform a top-$k$ nearest-neighbor search within the FAISS index. Candidates with a cosine similarity falling below a predefined baseline threshold are immediately discarded. This step acts as a rapid filter, isolating a diverse subset of lexically relevant matches for deeper analysis.

\textbf{Stage 2: Distortion-Aware Fine Matching.} Different transformations shatter distinct code aspects. For instance, control-flow flattening severely alters the structural layout, whereas standard optimization modifies local data dependencies. Consequently, relying on a uniform similarity metric is fundamentally insufficient. Instead, for a given target line $l$ and a retrieved candidate $c$, we predict the dominant distortion type of $l$. Based on this type, we assign a specific weight vector $w$. The final Dynamic Semantic Score (DSS) is computed by linearly aggregating the four dimensions of the previously defined Decompilation Semantic Signature (i.e., syntax features, variable dependencies, control flow, and contextual role), denoted as $\operatorname{sim}_i(l,c)$ for $i\in\{1,2,3,4\}$:
\begin{equation}
  \operatorname{DSS}(l,c)=\sum_{i=1}^{4} w_i\cdot\operatorname{sim}_i(l,c).
\end{equation}
This distortion-aware weighting ensures, for example, that control-flow similarity is heavily prioritized for obfuscated lines, whereas syntax-level similarity is favored for lightly optimized lines.

\textbf{Stage 3: Context-Bounded Annotation.} Considering the limited input token budget of LLMs and the risk of ``context poisoning'' from irrelevant hints, we enforce strict allocation bounds. First, a candidate is only accepted if its DSS strictly exceeds a high-confidence quality threshold. Second, to maintain a balanced prompt, we impose a hard cap on the number of injected comments per distortion type (e.g., retaining at most two suggestions). The highest-scoring candidates that satisfy these dual constraints are ultimately appended to the target lines as inline semantic guidance. By executing this pipeline, \tool successfully mitigates severe obfuscation by precisely injecting previously observed semantic patterns.

\subsection{Structure-Oriented Repair via Control Flow Prediction}
\label{sec:structure-repair}

Compiler optimization and obfuscation, such as loop unrolling, conditional merging, and control-flow flattening (CFF), often introduce severe structural distortions in decompiled outputs. Flattened control flow eliminates conventional nesting, merges disjoint branches, and replaces structured constructs with opaque state-machine logic. Consequently, directly recovering the original source structure is extremely challenging, as the hierarchical organization of loops, conditionals, and returns is no longer explicitly observable.

In real-world closed-source analysis, the exact combination of these transformations is essentially a black box. Instead of attempting to reconstruct complete structured code end-to-end, which often causes vanilla LLMs to suffer from severe context dispersion, we explicitly decouple structural reasoning from semantic generation. We introduce a control-flow prediction module that learns to infer the source-level control skeleton from distorted decompiled inputs. The key intuition is that, even when the surface syntax is flattened, subtle lexical and data-flow cues remain (e.g., recurring comparison operators, jump patterns, or state variables) that implicitly encode the original logic layout. By training on paired decompiled--source functions across diverse optimization and obfuscation transformations, the model learns to decode these cues into a compact, canonical sequence summarizing the original control flow.

\textbf{Prediction Objective.} Given a decompiled function that has been distorted by compiler optimization or code obfuscation, the module predicts a compact, linearized control-flow sequence that summarizes its source-level control skeleton. The output is a sequence of control keywords and structural tokens (e.g., \texttt{if (cond)}, \texttt{else}, \texttt{for (...)}, \texttt{\{}, \texttt{\}}) stripped of all semantic business logic, explicitly representing the pure hierarchical topology of the original source code.

\textbf{Source of Supervision.} The model is trained using decompiled code generated from both compiler optimization (O1--O3) and obfuscation schemes (e.g., CFF, BCF, SUB, SPLIT). Each training sample is paired with its corresponding ground-truth control flow sequence extracted from the original source function, enabling the model to learn transformation-invariant structural cues.

\textbf{Integration into the Recovery Pipeline.} During inference, the predicted control-flow sequence is prepended to the LLM prompt in a standardized template. This scaffold acts as an immutable structural prior. It serves as a rigid guiding constraint that forces the LLM to synthesize the semantically grounded blocks strictly within the correct architectural layout. As illustrated in Figure~\ref{fig:workflow}(b), the prediction provides an explicit control skeleton that compensates for flattened or reordered control flow in the decompiled input.

\textbf{Generalization Across Distortions.} Unlike rule-based reconstruction or heuristic graph recovery, our approach learns transformation invariant structural correlations directly from data. Once trained, the predictor can infer plausible source-level skeletons for unseen obfuscation schemes or optimization patterns, offering a robust, architecture-agnostic prior for downstream LLM recovery. This synergy between symbolic structure prediction and generative refinement bridges the gap between disordered low-level control and interpretable high-level logic, substantially improving structural and semantic fidelity in the final output.

\subsection{Lexical, Stylistic, and Semantic Finalization}
\label{sec:finalization}

After guided semantic recovery and structure-oriented repair, \tool performs a finalization stage that transforms intermediate outputs into source-like, human-readable code. While prior work such as \fide~\cite{zhou2025fidelitygpt} employs fidelity-oriented prompting for basic semantic preservation, \tool elevates this process into a \textbf{prior-constrained joint deduction}. We design prompts that explicitly anchor the LLM to the semantic and structural priors established in our earlier phases, guiding it to refine identifiers, types, low-level artifacts, and stylistic conventions while strictly preserving the recovered program logic.

The finalization phase addresses four complementary dimensions:
\begin{itemize}[leftmargin=*]
  \item \textbf{Lexical Finalization:} Recover meaningful variable and parameter names based on context and usage; recover accurate type annotations and symbolic macro references where applicable.
  \item \textbf{Syntactic Cleanup:} Eliminate redundant or synthetic constructs such as dead branches, empty control paths, or synthetic labels; reduce unnecessary nesting and simplify expressions.
  \item \textbf{Semantic Clarification:} Normalize magic constants, recover idiomatic error handling patterns, and clean up compiler-injected stubs or low-level operations irrelevant to the original logic.
  \item \textbf{Stylistic Harmonization:} Reformat indentation, modularize code layout, and ensure stylistic consistency---thereby aligning the final output with human-like readability and maintainability.
\end{itemize}

Technically, this stage leverages prompt-guided joint refinement. Rather than applying isolated post-processing heuristics or relying on unconstrained LLM guesswork, the model is explicitly instructed to reason over lexical, syntactic, and semantic cues simultaneously. By seamlessly combining the injected priors from previous recovery phases (i.e., the semantic suggestions and the control-flow skeleton) with our fidelity-oriented prompt templates, \tool produces outputs that are structurally coherent, semantically aligned, and visually readable.

This finalization phase completes the end-to-end recovery workflow, ensuring that all three dimensions, lexical, syntactic, and semantic, are jointly refined. Through this approach, the recovered code achieves higher fidelity to the original source in both structure and semantic intent, substantially improving human readability and source-level alignment for downstream analysis.

\section{Evaluation}
\label{sec:evaluation}

This section evaluates \tool from complementary perspectives.
\begin{itemize}[leftmargin=*]
  \item \textbf{RQ1:} How closely does the recovered code resemble the original source code?
  \item \textbf{RQ2:} How accurately can the original functions be retrieved from the recovered code?
  \item \textbf{RQ3:} How do different recovery stages contribute to semantic restoration under compiler optimizations and obfuscations? (Ablation study)
\end{itemize}

\subsection{Experiment Setup}

\textbf{Dataset.} We evaluate \tool on a diverse corpus of real-world C/C++ programs from established binary analysis benchmarks~\cite{marcelli2022machine} (e.g., binutils, coreutils, openssl, ImageMagick). Binaries (x86\_64) are decompiled using IDA Pro 7.5. By extracting functions compiled under three optimization levels (O1--O3) and four obfuscations (BCF, CFF, SPLIT, SUB) provided in the benchmark, we curated over 80,000 decompiled-source function pairs. From this, the Semantic Distortion Database was populated using 1,400 transformed functions (yielding 25,000+ suggestions). Our Test Set comprises 700 strictly isolated functions, meticulously excluded from all model training and database construction stages to prevent data leakage.

\textbf{Implementation Details.} \tool is implemented using Hugging Face Transformers and Tree-sitter for AST parsing. For structural prediction, we fine-tune CodeT5-base; its encoder-decoder architecture naturally suits the translation of flattened tokens into hierarchical control skeletons. For semantic retrieval, we select GraphCodeBERT-base. Its explicit pre-training on data-flow graphs aligns with our Decompilation Semantic Signature (DSS), capturing deep variable dependencies under severe distortion better than standard AST encoders.

Empirically optimized retrieval parameters include a FAISS pool size of $k=10$, a DSS threshold of $\tau=0.5$, and a cap of 2 suggestions per distortion type, balancing semantic recall with the prevention of context poisoning.

All generative tasks use the DeepSeek-Reasoner API~\cite{deepseek} at a temperature of 0.5. To ensure fair baseline comparisons, the maximum prompt length is bounded to 8,192 tokens, and baseline prompt lengths are padded or truncated to align tightly with \tool's inputs.

\subsection{Metrics}

We evaluate the effectiveness of \tool using two complementary paradigms: multi-level code similarity and retrieval-based semantic identifiability.

\textbf{Code Similarity Metrics.} We quantify how closely the recovered function aligns with the original source across five normalized dimensions: (1) \textbf{Interface Similarity}: Levenshtein distance of function names, return types, and parameters; (2) \textbf{Structural Similarity}: Edit distance between Tree-sitter parsed Abstract Syntax Trees (ASTs); (3) \textbf{Control Flow Similarity}: Sequence alignment of extracted control-flow tokens; (4) \textbf{Halstead Similarity}~\cite{thirumalai2017assessment}: Variance in cognitive complexity volume; and (5) \textbf{Token Edit Similarity}: Normalized edit distance over lexical streams. \textbf{Note on Metric Orthogonality:} While AST and CFG similarities may appear conceptually overlapping, they capture fundamentally distinct signals under obfuscation. For instance, CFF completely dismantles the AST hierarchy, rendering AST similarity near zero, yet preserving the sequence of isolated control tokens (CFG similarity) remains a robust measure of skeleton recovery. Exact mathematical formulations for all metrics are detailed in~\cite{ReSource}.

\textbf{Retrieval-Based Function Detection.} Aggressive transformations render traditional token-overlap metrics (e.g., BLEU~\cite{papineni2002bleu}) insufficient, as they severely penalize structurally divergent but functionally equivalent code. Therefore, rather than demanding exact textual equivalence, we evaluate recovery quality from the perspective of semantic identifiability. We assess whether a recovered function preserves sufficient semantic signals to successfully retrieve its ground-truth source from a massive candidate corpus containing 10,853 distinct source functions. We embed both recovered and source functions using GraphCodeBERT, which leverages data-flow edge modeling to capture deeper semantics beyond lexical overlap. To prevent trivial identifier leakage, all function names are normalized to \texttt{FUNC} and comments are stripped prior to embedding. We report the Top-$k$ ($k=1,3,5$) retrieval accuracy, providing a robust proxy for the code's utility in downstream security tasks like vulnerability matching.

\subsection{Baselines}

To contextualize our results, we compare \tool with several baselines and ablation variants:
\begin{itemize}[leftmargin=*]
  \item \degpt~\cite{hu2024degpt}: A state-of-the-art decompilation optimization framework that uses the DeepSeek-R1 LLM API. It focuses on variable renaming, code structure simplification, and comment generation. We exclude comment generation in our comparison.
  \item \fide~\cite{zhou2025fidelitygpt}: A framework for decompilation repair, utilizing the DeepSeek-R1 LLM API. We exclude comment generation and compare the corrected output generated during its correction phase.
  \item \llm~\cite{tan2024llm4decompile}: Refers to the LLM4Decompile series, trained on both source and assembly code with support for O1--O3 optimization levels.
  \item \ReSourcec: Disable semantic distortion suggestion and only rely on control-flow and lexical finalization.
  \item \ReSourceg: Remove control-flow-guided structure recovery and retain only semantic recovery and finalization.
  \item \ReSourcef: Retain only the finalization stage and remove both semantic and structural suggestion.
\end{itemize}
These baselines allow us to evaluate the individual contributions of each stage and benchmark \tool against current state-of-the-art approaches.

\subsection{RQ1: Code Similarity After Recovery}
\label{sec:rq1}

Table~\ref{tab:rq1_similarity} presents the multi-dimensional code similarity results. \tool consistently achieves the highest overall similarity (average 0.66), outperforming \fide (0.62), \llm (0.61), and \degpt (0.58) across all transformation types. Notably, \tool's performance remains remarkably stable even under heavy compiler optimization (O3) and adversarial obfuscation (e.g., CFF), which severely degrade existing approaches.

\textbf{Baseline Overfitting and Partial Repair.} A closer inspection reveals stark differences in how methods handle adversarial versus benign transformations. As highlighted in Table~\ref{tab:rq1_similarity}, \llm manages the highest average similarity under pure optimizations (Opt-AVG: 0.67), heavily dominating surface-level Interface Similarity. This demonstrates that its large-scale fine-tuning successfully memorizes standard compiler patterns. However, its performance catastrophically collapses under obfuscation (Obf-AVG: 0.58), exposing severe overfitting to benign transformations. Meanwhile, although \fide's consistency-checking heuristics allow it to maintain competitive Control Flow Similarity under certain obfuscations, it fundamentally fails to resolve holistic structural distortions, lagging significantly in Structural and Token Edit metrics under severe settings like O3 and CFF.

\textbf{\tool's Transformation-Agnostic Robustness.} In stark contrast, \tool demonstrates true transformation-agnostic robustness. While it does not overfit to the memorized formatting of O1/O2, it maintains an impressive and stable Obf-AVG of 0.67. It consistently captures the highest Structural, Halstead, and Token Editing similarities across O3 and all obfuscation settings. These distinct gains prove that \tool's combination of structural skeleton prediction and semantic guidance successfully recovers complex, high-level logic that fine-tuned baselines completely miss. Rather than performing superficial lexical patches, \tool's multi-stage recovery produces representations that genuinely reflect the original source architecture, securing the highest overall average similarity (0.66).

\begin{table*}[t]
\centering
\caption{Similarity metrics across optimization and obfuscation. The table is grouped to highlight performance degradation under obfuscation. Best results are in \textbf{bold}, second-best are \underline{underlined}.}
\label{tab:rq1_similarity}
\tiny
\setlength{\tabcolsep}{3pt}
\begin{tabular}{cl cccc ccccc c}
\toprule
\multirow{2}{*}{\textbf{Metric}} & \multirow{2}{*}{\textbf{Approach}} & \multicolumn{4}{c}{\textbf{Optimization}} & \multicolumn{5}{c}{\textbf{Obfuscation}} & \multirow{2}{*}{\textbf{Overall}} \\
\cmidrule(lr){3-6}\cmidrule(lr){7-11}
& & \textbf{O1} & \textbf{O2} & \textbf{O3} & \textbf{Opt-AVG} & \textbf{BCF} & \textbf{CFF} & \textbf{Split} & \textbf{Sub} & \textbf{Obf-AVG} & \textbf{AVG} \\
\midrule
\multirow{4}{*}{Interface} & \degpt & 0.64&0.64&0.64&0.64&0.62&0.64&0.64&0.65&0.64&0.64\\
& \fide & 0.69&0.69&0.68&0.69&0.68&0.70&0.68&0.69&0.69&0.69\\
& \llm & \textbf{0.78}&\textbf{0.77}&\textbf{0.78}&\textbf{0.78}&\textbf{0.77}&\textbf{0.77}&\textbf{0.77}&\textbf{0.78}&\textbf{0.77}&\textbf{0.78}\\
& \tool & \underline{0.69}&\underline{0.71}&\underline{0.72}&\underline{0.71}&\underline{0.70}&\underline{0.71}&\underline{0.71}&\underline{0.74}&\underline{0.72}&\underline{0.71}\\
\midrule
\multirow{4}{*}{Structural} & \degpt & 0.50&0.48&0.44&0.47&0.49&0.44&0.51&0.52&0.49&0.48\\
& \fide & 0.52&0.51&0.46&0.50&0.54&0.49&0.54&0.57&0.54&0.52\\
& \llm & \textbf{0.60}&\textbf{0.59}&\underline{0.53}&\textbf{0.57}&0.48&0.42&0.48&\underline{0.58}&0.49&0.52\\
& \tool & \underline{0.55}&\underline{0.54}&\textbf{0.55}&\underline{0.55}&\textbf{0.57}&\textbf{0.53}&\textbf{0.56}&\textbf{0.58}&\textbf{0.56}&\textbf{0.55}\\
\midrule
\multirow{4}{*}{Control Flow} & \degpt & 0.73&\textbf{0.72}&\underline{0.69}&\textbf{0.71}&\underline{0.74}&0.59&\underline{0.75}&\textbf{0.79}&\underline{0.72}&0.70\\
& \fide & 0.72&0.70&0.65&0.69&\textbf{0.77}&\textbf{0.67}&\textbf{0.77}&\textbf{0.79}&\textbf{0.75}&\textbf{0.72}\\
& \llm & 0.72&\textbf{0.72}&0.65&\underline{0.70}&0.51&0.39&0.48&0.69&0.52&0.58\\
& \tool & \textbf{0.74}&0.70&\textbf{0.70}&\textbf{0.71}&\underline{0.76}&\textbf{0.67}&0.72&\underline{0.76}&0.73&\textbf{0.72}\\
\midrule
\multirow{4}{*}{Halstead} & \degpt & 0.71&0.68&0.61&0.67&0.69&0.55&0.73&0.75&0.68&0.66\\
& \fide & 0.71&0.69&0.60&0.67&0.71&0.64&0.71&0.77&0.71&0.69\\
& \llm & \textbf{0.79}&\textbf{0.79}&\underline{0.69}&\textbf{0.76}&0.65&0.55&0.64&0.75&0.65&0.69\\
& \tool & \underline{0.77}&\underline{0.76}&\textbf{0.76}&\textbf{0.76}&\textbf{0.76}&\textbf{0.69}&\textbf{0.77}&\textbf{0.80}&\textbf{0.76}&\textbf{0.75}\\
\midrule
\multirow{4}{*}{Token Editing} & \degpt & 0.45&0.44&0.41&0.43&0.45&0.37&0.47&0.48&0.44&0.43\\
& \fide & 0.50&0.50&0.44&0.48&0.53&0.46&0.52&0.55&0.52&0.50\\
& \llm & \textbf{0.57}&\textbf{0.55}&0.49&\textbf{0.54}&0.48&0.39&0.49&0.57&0.48&0.49\\
& \tool & \underline{0.55}&\underline{0.54}&\textbf{0.54}&\textbf{0.54}&\textbf{0.57}&\textbf{0.53}&\textbf{0.57}&\textbf{0.59}&\textbf{0.57}&\textbf{0.55}\\
\midrule
\multirow{4}{*}{Overall AVG} & \degpt & 0.61&0.59&0.56&0.59&0.60&0.52&0.62&0.64&0.60&0.58\\
& \fide & 0.63&0.62&0.57&0.61&0.64&0.59&0.64&0.67&0.64&0.62\\
& \llm & \textbf{0.69}&\textbf{0.69}&0.63&\textbf{0.67}&0.58&0.51&0.57&0.67&0.58&0.61\\
& \tool & \underline{0.66}&\underline{0.65}&\textbf{0.65}&\underline{0.65}&\textbf{0.67}&\textbf{0.63}&\textbf{0.67}&\textbf{0.69}&\textbf{0.67}&\textbf{0.66}\\
\bottomrule
\end{tabular}
\end{table*}

\begin{tcolorbox}[breakable,colback=gray!15,colframe=gray!15,boxrule=0pt]
\textbf{Answer to RQ1:} Benefiting from its unified, decoupled framework, \tool achieves the most stable and robust recovery across the entire spectrum of transformations. While baseline models may peak under specific benign optimizations, \tool actively resists overfitting and maintains high fidelity even under severe structural destruction, securing the highest overall performance.

\textbf{Insight 1.1: Obfuscation introduces deeper semantic distortion than optimization.} Our data shows that while standard optimizations preserve basic logical flow, adversarial obfuscations shatter the control skeleton, causing a catastrophic drop in baseline performance.

\textbf{Insight 1.2: Optimization-only fine-tuning severely overfits.} Models relying strictly on compilation corpora (e.g., \llm) memorize superficial token arrangements but completely fail to learn transformation-agnostic semantics, severely limiting their real-world generalization.
\end{tcolorbox}

\subsection{RQ2: Function Retrieval Accuracy}
\label{sec:rq2}

Table~\ref{tab:rq2_retrieval} details the Top-$k$ ($k=1,3,5$) retrieval accuracy. \tool achieves a remarkable average Top-5 accuracy of 83\%, demonstrating a clear semantic advantage over \fide (77\%), \degpt (65\%), and \llm (61\%). This confirms that \tool successfully restores representations that align with the developer's original intent, even under aggressive transformations.

\textbf{Baseline Vulnerabilities vs. \tool's Robustness.} A closer analysis reveals the limitations of relying solely on end-to-end fine-tuning. \llm manages a competitive Top-1 retrieval under pure optimizations (Opt-AVG: 62\%), even outperforming \tool under O2. However, this localized peak exposes its reliance on memorizing benign artifacts; once subjected to structural obfuscation, its Top-1 accuracy catastrophically collapses (Obf-AVG: 43\%). Similarly, while \fide achieves a marginally higher Top-5 accuracy under localized substitution (Sub: 85\%), it degrades sharply under complex control-flow shifts like Split. In stark contrast, \tool maintains resilient identifiability, achieving a Top-1 Opt-AVG of 67\% and an even more impressive Obf-AVG of 70\%.

\textbf{The Disconnect Between Similarity and Fidelity.} Under O1 optimization, \llm achieves a deceptively high average token similarity (0.69, slightly edging out \tool's 0.66 in RQ1), yet its actual retrieval accuracy is strictly lower than \tool's. This proves that high token overlap often stems from formatting or superficial syntax matching, whereas retrieval accuracy rigorously evaluates global semantic consistency.

\begin{table*}[t]
\centering
\caption{Function retrieval accuracy across optimization and obfuscation. The table is grouped to highlight performance degradation under obfuscation. Best results are in \textbf{bold}, second-best are \underline{underlined}.}
\label{tab:rq2_retrieval}
\tiny\setlength{\tabcolsep}{3pt}
\begin{tabular}{cl cccc ccccc c}
\toprule
\multirow{2}{*}{\textbf{Metric}} & \multirow{2}{*}{\textbf{Approach}} & \multicolumn{4}{c}{\textbf{Optimization}} & \multicolumn{5}{c}{\textbf{Obfuscation}} & \multirow{2}{*}{\textbf{Overall}}\\
\cmidrule(lr){3-6}\cmidrule(lr){7-11}
& & \textbf{O1}&\textbf{O2}&\textbf{O3}&\textbf{Opt-AVG}&\textbf{BCF}&\textbf{CFF}&\textbf{Split}&\textbf{Sub}&\textbf{Obf-AVG}&\textbf{AVG}\\
\midrule
\multirow{4}{*}{Top-1 (\%)} & \degpt &57&52&41&50&53&30&57&61&50&50\\
& \fide &62&60&54&59&65&60&59&\underline{72}&64&62\\
& \llm &\underline{69}&\textbf{66}&51&62&38&27&44&62&43&51\\
& \tool &\textbf{70}&\underline{64}&\textbf{66}&\textbf{67}&\textbf{68}&\textbf{62}&\textbf{76}&\textbf{73}&\textbf{70}&\textbf{68}\\
\midrule
\multirow{4}{*}{Top-3 (\%)} & \degpt &64&67&50&60&67&44&66&68&61&61\\
& \fide &\textbf{79}&69&67&72&74&72&70&78&74&73\\
& \llm &78&\textbf{77}&60&72&47&30&48&71&49&59\\
& \tool &\textbf{79}&73&\textbf{81}&\textbf{78}&\textbf{80}&\textbf{78}&\textbf{84}&\textbf{82}&\textbf{81}&\textbf{80}\\
\midrule
\multirow{4}{*}{Top-5 (\%)} & \degpt &72&68&57&66&71&46&69&71&64&65\\
& \fide &80&74&70&75&78&76&73&\textbf{85}&78&77\\
& \llm &79&\textbf{80}&63&74&48&34&49&72&51&61\\
& \tool &\textbf{82}&79&\textbf{83}&\textbf{81}&\textbf{83}&\textbf{84}&\textbf{85}&84&\textbf{84}&\textbf{83}\\
\bottomrule
\end{tabular}
\end{table*}

\begin{tcolorbox}[breakable,colback=gray!15,colframe=gray!15,boxrule=0pt]
\textbf{Answer to RQ2:} \tool yields substantially higher retrieval accuracy than all baselines, achieving an impressive 83\% average Top-5 accuracy. This demonstrates its superior capability to restore genuine semantic fidelity that aligns with the developer's original intent.

\textbf{Insight 2.1: Lexical similarity does not imply semantic identifiability.} High token overlap often masks severe semantic inconsistencies. Our findings confirm that source-level retrieval is a far more rigorous and accurate proxy for evaluating decompilation success than shallow metrics like BLEU or edit distance.
\end{tcolorbox}

\subsection{RQ3: Effect of Each Phase (Ablation)}
\label{sec:rq3}

To understand the individual contributions of our multi-stage architecture, Table~\ref{tab:rq3_retrieval} presents an ablation study isolating semantic suggestion (\ReSourceg), structural prediction (\ReSourcec), and the baseline lexical finalization (\ReSourcef).

\begin{table*}[t]
\centering
\caption{Ablation study: Function retrieval accuracy across optimization and obfuscation. Best results among the isolated variants are highlighted in \textbf{bold}, and second-best are \underline{underlined}.}
\label{tab:rq3_retrieval}
\tiny\setlength{\tabcolsep}{3pt}
\begin{tabular}{cl cccc ccccc c}
\toprule
\multirow{2}{*}{\textbf{Metric}} & \multirow{2}{*}{\textbf{Approach}} & \multicolumn{4}{c}{\textbf{Optimization}} & \multicolumn{5}{c}{\textbf{Obfuscation}} & \multirow{2}{*}{\textbf{Overall}}\\
\cmidrule(lr){3-6}\cmidrule(lr){7-11}
& & \textbf{O1}&\textbf{O2}&\textbf{O3}&\textbf{Opt-AVG}&\textbf{BCF}&\textbf{CFF}&\textbf{Split}&\textbf{Sub}&\textbf{Obf-AVG}&\textbf{AVG}\\
\midrule
\multirow{3}{*}{Top-1 (\%)} & \ReSourceg &68&67&\textbf{66}&\textbf{67}&66&\underline{71}&68&\underline{71}&\underline{69}&\textbf{68}\\
& \ReSourcec &\textbf{71}&\textbf{68}&\underline{58}&\underline{66}&\underline{67}&\textbf{74}&\underline{67}&\textbf{74}&\textbf{71}&\textbf{68}\\
& \ReSourcef &\underline{64}&\underline{61}&56&60&\textbf{68}&52&\textbf{71}&68&65&63\\
\midrule
\multirow{3}{*}{Top-3 (\%)} & \ReSourceg &76&76&\textbf{74}&75&77&78&\textbf{81}&\textbf{80}&\textbf{79}&\underline{77}\\
& \ReSourcec &\textbf{82}&\textbf{81}&\underline{69}&\textbf{77}&\underline{77}&\textbf{81}&\underline{77}&\underline{78}&\underline{78}&\textbf{78}\\
& \ReSourcef &\underline{75}&70&64&70&\textbf{80}&61&80&78&75&73\\
\midrule
\multirow{3}{*}{Top-5 (\%)} & \ReSourceg &79&78&\textbf{79}&79&79&81&\textbf{85}&\underline{83}&\underline{82}&\textbf{81}\\
& \ReSourcec &\textbf{85}&\textbf{82}&\underline{70}&\textbf{79}&\textbf{81}&\textbf{86}&\underline{82}&\textbf{84}&\textbf{83}&\textbf{81}\\
& \ReSourcef &\underline{77}&75&68&73&\textbf{81}&67&81&81&78&76\\
\bottomrule
\end{tabular}
\end{table*}

\textbf{Orthogonal Strengths.} The results reveal that the semantic and structural modules target distinct, non-overlapping dimensions of the binary-to-source gap. Specifically, \ReSourcec (which strictly adds structural prediction) exhibits its most pronounced gains under heavy control-flow obfuscation, achieving an impressive 86\% Top-5 accuracy under CFF. This proves its necessity for un-flattening distorted logic. However, its performance drops noticeably under heavy optimization (O3 Top-5: 70\%), where structural hierarchy is preserved but semantic cues are aggressively stripped. Conversely, \ReSourceg (which provides database-driven semantic guidance) demonstrates dominant stability across O3 optimizations (Top-5: 79\%), effectively restoring high-level intent. Unsurprisingly, the fully ablated variant \ReSourcef---relying solely on vanilla LLM-driven lexical refinement---exhibits the poorest global performance (Top-5 AVG: 76\%), confirming that superficial token patching cannot resolve deep semantic shifts.

\textbf{Synergistic Full Pipeline.} The complete \tool pipeline achieves the optimal balance. As established in RQ2, the full \tool architecture achieves an overall Top-5 average of 83\%, strictly outperforming the best isolated variants here (which plateau at 81\%). This synergy confirms that semantic suggestion and structural prediction are highly complementary rather than redundant: Phase 2 reconstructs the architectural skeleton, while Phase 1 injects the semantic flesh. Case studies demonstrating how these stages collaborate are available in our replication package~\cite{ReSource}.

\begin{tcolorbox}[breakable,colback=gray!15,colframe=gray!15,boxrule=0pt]
\textbf{Answer to RQ3:} The complete \tool pipeline (83\% Top-5 AVG) significantly outperforms its isolated variants (max 81\%). Removing either the semantic database or the structural predictor causes a notable drop in accuracy, confirming that surface-level generation alone (\ReSourcef at 76\%) is fundamentally insufficient.

\textbf{Insight 3.1: Semantic and structural recovery are orthogonal yet synergistic.} While each phase targets distinct aspects of the semantic gap (e.g., structural prediction uniquely resolves control-flow flattening, while semantic guidance restores stripped intent under O3), their combination is strictly necessary for robust source recovery.
\end{tcolorbox}

\section{Discussion and Limitations}
\label{sec:discussion}

\subsection{Discussion and Implications}

\textbf{Beyond Surface-Level Similarity.} Our empirical results highlight a consistent discrepancy between token-level similarity metrics (e.g., BLEU) and actual semantic recovery (retrieval accuracy). Under aggressive obfuscation, sequence-to-sequence models may produce visually plausible code that fundamentally fails to preserve the original logic. By explicitly decoupling semantic and structural constraints, \tool demonstrates that robust decompilation requires guiding the LLM with empirical priors, rather than relying solely on lexical pattern matching.

\textbf{Facilitating Source-Level Security Analysis.} A practical implication of \tool is its potential to ease the burden of binary analysis. By lifting obfuscated binaries into readable, high-fidelity source representations, \tool helps bridge the gap between hostile binary environments and the mature ecosystem of source-level static analysis tools (e.g., CodeQL, Fortify). This allows security analysts to inspect logic and vulnerabilities in a more intuitive domain.

\textbf{The Value of Modular Recovery.} The ablation study (\S\ref{sec:rq3}) suggests that semantic hallucinations and structural collapse are distinct failure modes that benefit from targeted treatments. The synergistic performance of \tool's pipeline indicates that modular, neuro-symbolic architectures---where neural generation is constrained by symbolic structural skeletons---offer a promising direction for future reverse engineering frameworks.

\subsection{Limitations and Future Work}

While \tool improves the robustness of source code recovery, several limitations remain to be addressed in future work:

\textbf{Defining Success via Semantic Fidelity instead of Execution.} Verifying formal functional equivalence (e.g., via unit testing or symbolic execution) for isolated functions is often an ill-posed problem due to the absence of global context (e.g., missing header definitions, global variables, or external function prototypes). Consequently, \tool prioritizes semantic fidelity by ensuring that the recovered code preserves the original logic and security-critical properties as a proxy for correctness.

\textbf{Scope of Semantic Database and Generalization.} While the Semantic Distortion Database provides a principled way to reason about optimizations, its current coverage is primarily centered on transformation patterns from mainstream compilers (e.g., GCC, Clang) and common obfuscators. Its performance on highly customized or proprietary obfuscation schemes remains untested. As a data-driven framework, \tool's generalization capability is inherently tied to the diversity of its underlying database.

\textbf{Interprocedural Context and Global State.} Currently, \tool operates at the function level. Since complete program context is often unavailable during standalone function recovery, the framework might struggle with functions whose semantics are deeply coupled with external state or complex cross-function data flows. Extending \tool toward interprocedural reasoning and global type inference is a key future direction.

\section{Related Work}
\label{sec:RelatedWork}

\textbf{From Binary Analysis to Decompiled Understanding.} Early research on binary similarity and recovery focused on low-level representations such as instruction sequences, control-flow graphs, and intermediate representations (IR). Learning-based approaches and graph neural networks were used to extract transformation-invariant embeddings for robust binary understanding. For example, ImOpt~\cite{jiang2020similarity} normalizes instruction patterns to reduce compiler-induced redundancy; GTrans~\cite{zhang2024gtrans} models hierarchical structures in control-flow graphs to mitigate obfuscation-induced reshaping; VulHawk~\cite{luo2023vulhawk} transfers binary embeddings across architectures via microcode lifting; and UniASM~\cite{gu2025uniasm} unifies binary embeddings to achieve resilience across both optimization and obfuscation. Although effective for structural matching, these methods remain constrained by their low-level focus, lacking access to types, variables, and logical intent, and thus cannot recover readable, source-like representations.

\textbf{From Decompiled Code to Type and Identifier Recovery.} The advent of modern decompilers (e.g., Ghidra~\cite{ghidra}, IDA Pro~\cite{hex-rays}) shifted the focus toward decompiled pseudocode, offering a structured foundation for LLMs to reason over higher-level abstractions. Consequently, tools like ReSym~\cite{xie2024resym} and TypeForge~\cite{wang2025typeforge} integrate LLMs with static analysis to successfully infer stripped identifiers and reconstruct composite types. These works significantly improve surface readability, laying essential groundwork for more comprehensive semantic restoration.

\textbf{LLM-Based Decompilation and Code Refinement.} Recent advancements apply LLMs directly to end-to-end decompilation or post-processing. For instance, \llm~\cite{tan2024llm4decompile} fine-tunes models on assembly-to-C datasets to recover source code across standard optimizations (O1--O3). Concurrently, general-purpose frameworks like \degpt~\cite{hu2024degpt} and \fide~\cite{zhou2025fidelitygpt} employ prompt-guided strategies for variable renaming and structural simplification, achieving substantial gains in lexical and syntactic quality. Despite these significant advancements, preserving deep semantic fidelity and structural resilience under aggressive, adversarial obfuscation remains an open and complex challenge in the field.

\section{Conclusion}
\label{sec:Conclusion}

We presented \tool, a multi-phase LLM framework designed to bridge the severe semantic and structural gaps introduced by aggressive optimization and adversarial obfuscation. By fundamentally decoupling the recovery process, \tool neutralizes the semantic hallucinations and structural collapse that plague monolithic, end-to-end LLM approaches. Specifically, the synergistic integration of empirical priors (via the Semantic Distortion Database) and structural skeletons (via the control-flow predictor) enables \tool to achieve true transformation-agnostic robustness.

Extensive evaluations on a massive benchmark spanning diverse optimization levels and obfuscation schemes demonstrate that \tool attains an average source-level similarity of 0.66 and an 83\% Top-5 retrieval accuracy. Ultimately, \tool demonstrates that explicitly decoupling structural and semantic constraints significantly improves decompilation fidelity, offering a promising step toward reliable source-level security analysis.

\section{Data Availability Statement}

The core artifacts, which include the full implementation of \tool, the curated decompiled-source dataset, and the Semantic Distortion Database, are permanently archived on Zenodo via the following DOI: \url{https://doi.org/10.5281/zenodo.19212869}.

For convenient browsing, additional supplementary materials such as detailed prompt templates, complete technical details, extra experimental studies, and qualitative case studies are hosted on our project website~\cite{ReSource}.

\bibliographystyle{ACM-Reference-Format}
\bibliography{reference}

\end{document}